# Implicit Regression: Detecting Constants and Inverse Relationships with Bivariate Random Error

R. D. Wooten, K. Baah, J. D'Andrea
Department of Mathematics and Statistics
University of South Florida, Tampa

**Abstract**: Regression has been a statistical tool since 1805. In 2011, Wooten introduced Non-Response Analysis the founding theory in Implicit Regression where Implicit Regression treats the variables implicitly as co-dependent variables and not as an explicit function with dependent/independent variables as in standard regression. The motivation of this paper is to introduce methods of implicit regression to determine the constant nature of a variable ($x$ or $y$) or the interactive term ($xy$), and address inverse relationship among measured variables with random error present in both directions. The contribution of this research include an underlying theory to better address co-dependent relationship among measured variables with normal random error, and specifically, detecting constants and inverse relationships with bivariate random error. In this paper, we first used simulated data to compare this newly developed theory with standard statistical methods in order to validate such methodology. We then demonstrated how this theory can be used to verify Boyle's law using the data gathered by Boyle in 1662. Boyle's law states that pressure and volume of an ideal gas are inversely proportional. Thus, the product of pressure and volume are constant and subject to the natural random error that occurs in their measurements.

## 1. Introduction

Implicit Regression was developed by R. Wooten to better address co-dependent relationship among measured variables with normal random error. For the equation containing exactly two variables of interest,

$$g(x, y) = h(x, y|\theta),$$

where $g(x, y)$ is a fixed function with well-defined constant coefficients and $h(x, y)$ is defined in terms of the unknown coefficients, $\theta = \{\alpha_0, \alpha_1, \alpha_2, \ldots, \alpha_m\}$.

Given the terms are $\{1, x, y, xy\}$, then there are three rotations and one non-response model:

$$y = \alpha_0 + \alpha_1 x + \alpha_2 xy,$$
$$x = \alpha_0 + \alpha_1 y + \alpha_2 xy,$$
$$xy = \alpha_0 + \alpha_1 x + \alpha_2 y,$$
$$1 = \alpha_1 x + \alpha_2 y + \alpha_3 xy,$$

which can be used to analyze the nature the relationship that exist between the underlying variables.

## 2. Comparing Non-Response to Standard Regression for Univariate

In standard regression (Bulmer, 2003), we have that the subject response ($y$) is constant ($\beta = \mu$),
$$y = \beta$$
and that there is random error in the observed data,
$$y_i = \beta + \varepsilon_i.$$
where $E(\varepsilon) = 0$ and $V(\varepsilon) = \sigma_y^2$; and parameter estimate given by
$$\hat{\beta} = \frac{\sum y_i}{n} = \hat{\mu}_y.$$

However, using the non-response model we have that the subject response ($y$) is a non-zero constant ($\mu$), but instead of minimize the error, rather minimizes the percent error,
$$\frac{y - \mu}{\mu} = \alpha y - 1;$$
or equivalently, modeling
$$\alpha y = 1$$
where the random error that exist is related to the coefficient of variation,($CV$); the ratio of standard deviation to the mean over the mean alone
$$\alpha y_i = 1 + \omega_i,$$
where $E(\omega) = 0$ and $V(\omega) = CV_y^2 = \frac{\sigma_y^2}{\mu_y^2}$; and parameter estimate given by
$$\hat{\mu}_y = \frac{1}{\hat{\alpha}} = \frac{\sum y^2}{\sum y},$$
a self-weighting mean.

### 2.1 Simulation and Evaluation of Models

Consider the simulation defined by the relationships: $t \sim U(1,10)$, $x = \frac{200}{t}$ and $y = 20t$, where $x_i = x + N(0, \sigma^2)$ and $y_i = y + N(0, \sigma^2)$.

Note that in theory $xy = 4000$, with the observed error, the view through the kaleidoscope is of the form:
$$x_i y_i = (x + \delta_i)(y + \varepsilon_i) = 4000 + \delta_i y + \varepsilon_i x + \delta_i \varepsilon_i.$$
hence, we will consider $x$, $y$ and $xy$ to be the terms of interest.

For comparison, consider the standard models, $y = f(x)$ as equivalently $y = \frac{4000}{x}$ with error terms of the form:
$$y_i = \frac{4000}{x_i + \delta_i} + \varepsilon_i \approx \frac{4000}{x_i} + \varepsilon_i,$$

or even using Taylor expansion (Randall, 2011), about $a = 1$,
$$y_i = 4000[1 + (x - 1) + (x - 1)^2 \dots] + \varepsilon_i$$
Finally, reversing $x$ and $y$, $x = f(y)$ as equivalently $x = \frac{4000}{y}$ with error terms of the form:

$$x_i = \frac{4000}{y_i + \varepsilon_i} + \delta_i \approx \frac{4000}{y_i} + \delta_i.$$

Consider two random samples of size 50; one with a standard deviation $\sigma = 1$ and a second with a standard deviation $\sigma = 5$.

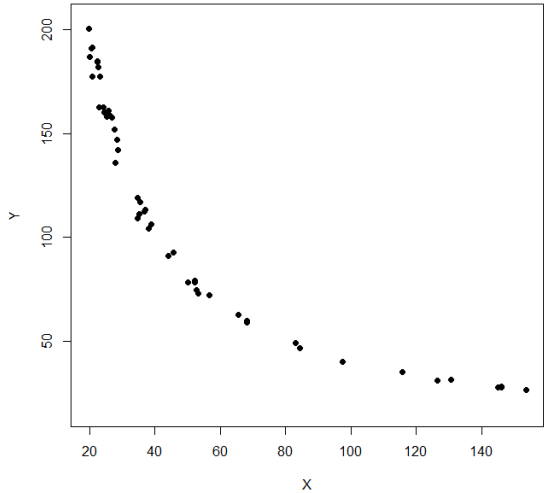
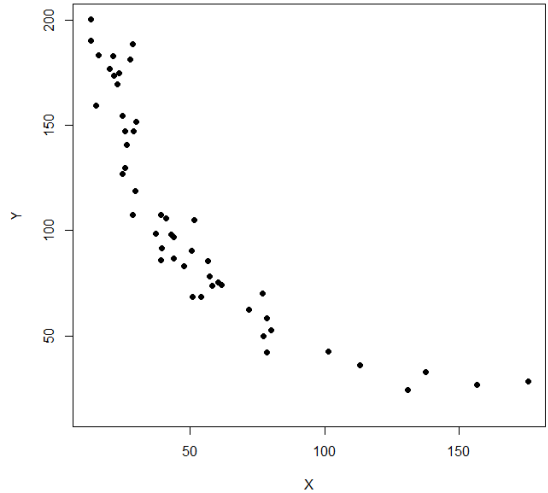

Fig. 1a                                   Fig. 1b

**Figure 1:** Scatter plot of simulated data with a) standard deviation of one ($\sigma = 1$) and b) standard deviation of five ($\sigma = 5$). In the Fig. 1a, the variance in both measures is set to one and it illustrates that the smaller the variance the easier the pattern is to detect. In Fig. 1b, the variance in both measures is set to twenty-five and the pattern becomes less clear and could be mistaken for a bounded quadratic.

When considering the constant of each term for the data with $\sigma = 5$: $x, y, xy$, we find that $xy$ has $R_{xy}^2 = 0.9989$, whereas $y$ is somewhat variant with $R_y^2 = 0.8041$ and $x$ is the most variant with $R_x^2 = 0.6452$. When considering the constant of each term for the data with $\sigma = 5$: $x, y, xy$, we find that $xy$ has $R_{xy}^2 = 0.9702$, whereas $y$ is somewhat variant with $R_y^2 = 0.8130$ and $x$ is the most variant with $R_x^2 = 0.6677$. That is, effects of the interaction term will be difficult to detect as $xy$ is approximately a constant. As illustrated in Figure 2(c) and Figure 3(c), the term $xy$ appears somewhat symmetric and concentrated around the mean. The variable $y$ appears to be rather uniformly distributed, Figure 2(b) and Figure 3(b); and its effect (if any) should be easier to detect. The variable $x$ in contrast has a clearly skewed distribution to the right and is rather variant, which means that $x$ as a function of $y$ will be easier to detect than $y$ as a function of $x$.

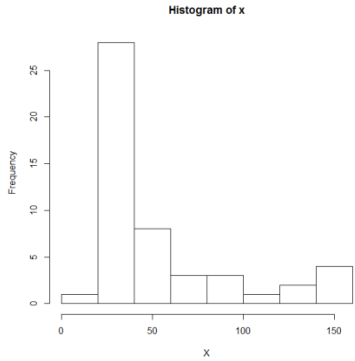 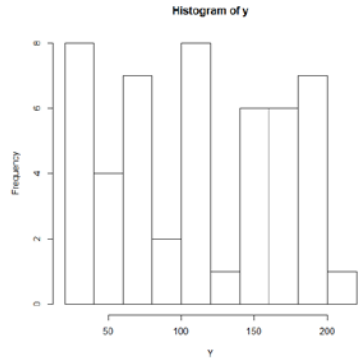 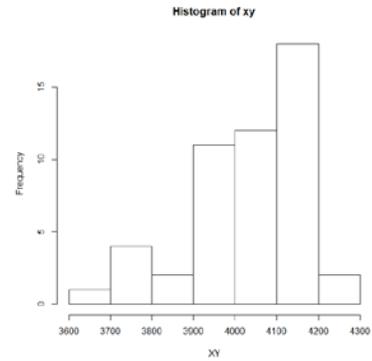

Fig. 2a            Fig. 2b            Fig. 2c

**Figure 2:** Histogram of: a) $x$ b) $y$ and c) $xy$ for simulated data with standard deviation ($\sigma = 1$). In Fig. 2a, the histogram depicts the first measure $x$ and shows the data distribution is skewed to the right. In Fig. 2b, the histogram depicts the second measure $y$ and shows the data distribution is approximately uniform. In Fig. 2c, the histogram depicts the product of the two measures $x$ and $y$; and shows the distribution is more symmetric than in the Figures 2a and 2b.

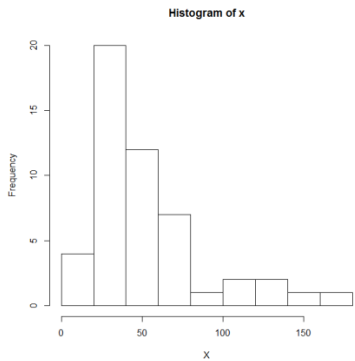 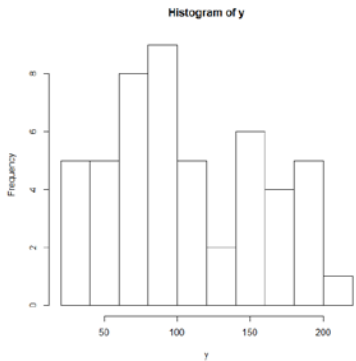 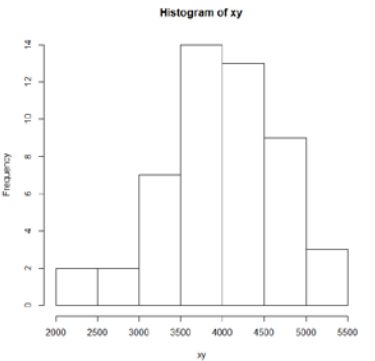

Fig. 3a            Fig. 3b            Fig. 3c

**Figure 3:** Histogram of: a) $x$ b) $y$ and c) $xy$ for simulated data with standard deviation ($\sigma = 5$). In Fig. 3a, the histogram depicts the first measure $x$ and shows the data distribution is skewed to the right. In Fig. 3b, the histogram depicts the second measure $y$ and shows the data distribution is approximately platykurtic. In Fig. 3c, the histogram depicts the product of the two measures $x$ and $y$; and shows that the distribution is approximately normal.

Comparing the various models created using implicit regression, we see that non-response analysis with all contributing terms ranks first in respect to coefficient of determination, estimated square error $y$, and non-response analysis with respect to only the interaction term ranks first in estimated square error of $x$.

Consider the simulation with standard deviation ($\sigma = 5$), Table 1. In the first rotation, $y = \alpha_0 + \alpha_1 x + \alpha_2 xy$, only the intercept and coefficient of $x$ are found to be significantly contributing; with a p-value of 0.856, the interaction term $xy$ is absorbed by the non-zero intercept. The reduced model is standard simple linear with $R^2 = 0.6981$. In the second rotation, $= \alpha_0 + \alpha_1 y + \alpha_2 xy$, once again only the intercept and coefficient of y are found to be significantly contributing, with a p-value of 0.113, the interaction term $xy$ is absorbed by the non-zero intercept. In the third rotation, $xy = \alpha_0 + \alpha_1 x + \alpha_2 y$, has the lowest $R^2$ value of 0.1316 for rotational analysis with $xy$ as a function of both x and y with an intercept, which is due to the fact that $xy$ is expected to be a constant of 4,000; hence the variables $x$ and $y$ only explain only 13% of the variance in $xy$. In univariate analysis, $R^2$ is a measure of the constant nature of the variable as previously outlined. Comparing these models using standard regression, the fourth model, $y = \beta_0 + \beta_1 \frac{1}{x}$ is unable to address the error in $x$; the relatively small error in $x$ is exacerbated when reciprocated resulting in $R^2 = 0.8208$. The fifth model, $y = \beta_0 + \beta_1 x + \beta_2 x^2$, is slightly better than the fourth model with an $R^2 = 0.8890$; however caution should be taken when extrapolating information. Continuing our comparison with non-response analysis, the sixth model, $1 = \alpha_1 x + \alpha_2 y + \alpha_3 xy$, is considered the best fitting model in terms of the coefficient of determination with $R^2 = 0.9852$. It also ranks first in terms of minimizing the standard error of $y$, and ranking fourth in minimizing the standard error of $x$. In the last model, $1 = \alpha_1 xy$, it is comparable to the rotational model $xy = \alpha_0$, with the same $R^2$ value is a good fit with similar estimates of the standard errors.

**Table 1**:  Coefficient of determination and estimations of Standard Errors

| Regression | Model | $\sigma = 1$ | | | $\sigma = 5$ | | |
|---|---|---|---|---|---|---|---|
| | | $R^2$ | $SE_y$ | $SE_x$ | $R^2$ | $SE_y$ | $SE_x$ |
| Rotational (Standard) | $y = \alpha_0 + \alpha_1 x + \alpha_2 xy$<br>$y = \alpha_0 + \alpha_1 x$ | 0.7575 (6.5) | 27.86 (4) | 24.80 (6) | 0.6981 (6.5) | 28.52 (4) | 24.80 (6) |
| Rotational (Standard) | $x = \alpha_0 + \alpha_1 y + \alpha_2 xy$<br>$x = \alpha_0 + \alpha_1 y$ | 0.7575 (6.5) | 32.02 (7) | 19.61 (6) | 0.6981 (6.5) | 34.13 (6) | 20.72 (5) |
| Rotational | $xy = \alpha_0 + \alpha_1 x + \alpha_2 y$<br>$xy = \alpha_0$ | 0.9989 (2) | 5.14 (3) | 1.42 (2) | 0.1316<br>0.9702 (2.5) | 32.75 (5) | 9.73 (2) |
| Standard | $y = \beta_0 + \beta_1 \frac{1}{x}$ | 0.9920 (4) | 5.07 (2) | 2.63 (4) | 0.8208 (5) | 21.97 (3) | 288.37 (7) |
| Standard | $y = \beta_0 + \beta_1 x + \beta_2 x^2$ | 0.9545 (5) | 12.19 (5) | 7.98 (5) | 0.8890 (4) | 17.48 (2) | 10.13 (3) |
| Non-Response | $1 = \alpha_1 x + \alpha_2 y + \alpha_3 xy$ | 0.9989 (2) | 4.94 (1) | 1.58 (3) | 0.9852 (1) | 16.40 (1) | 10.42 (4) |
| Non-Response | $1 = \alpha_1 xy$ | 0.9989 (2) | 5.16 (4) | 1.41 (1) | 0.9702 (2.5) | 35.25 (7) | 9.61 (1) |

The scatterplot of rotational models depicted in Figure 4 illustrates the lack of fit in the first two standard linear models and the apparent fit of the third model.

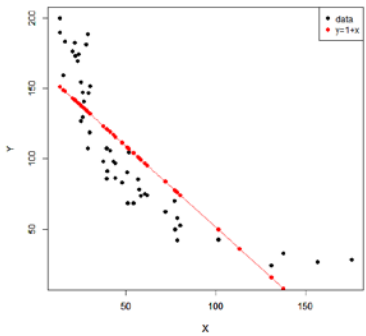 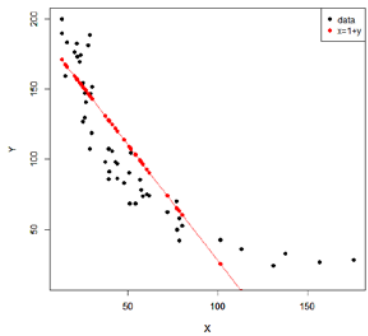 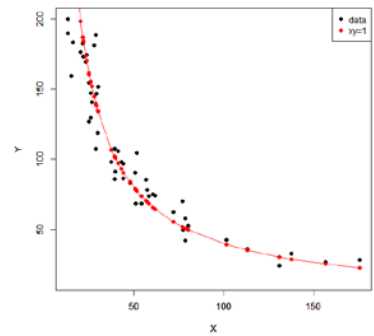

**Fig. 4a** **Fig. 4b** **Fig. 4c**

**Figure 4:** Scatter plot of rotational models; Fig. 4a illustrates the standard regression model $y = \alpha_0 + \alpha_1 x$; Fig. 4b illustrates the first rotation model $x = \alpha_0 + \alpha_1 y$; and Fig. 4c illustrates the non-response $xy = \alpha_0$. These models are the reduction of $y = \alpha_0 + \alpha_1 x + \alpha_2 xy$, $x = \alpha_0 + \alpha_1 y + \alpha_2 xy$, and $xy = \alpha_0 + \alpha_1 x + \alpha_2 y$ when only significant terms are included. The first two reduced models result in simple linear regression.

The scatterplots of standard regression models, Figure 5, illustrate the inability of the standard regression to address the random error in x; overestimating the value of y in the first model and underestimating the value of y in the second model.

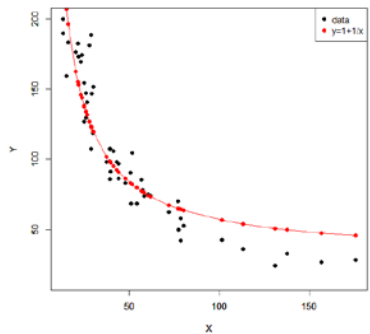 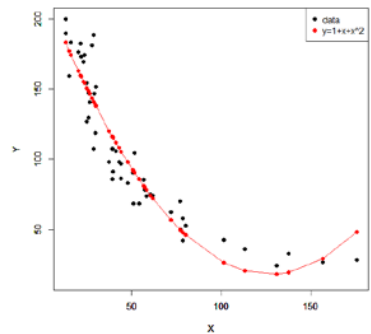

**Fig. 5a** **Fig. 5b**

**Figure 5:** Scatter plot of standard models; Fig. 5a illustrates the standard regression model $y = \beta_0 + \beta_1 \frac{1}{x}$; Fig 5b illustrates the standard regression model $y = \beta_0 + \beta_1 x + \beta_2 x^2$. These two figures demonstrate the weakness of standard regression methods in the detection of codependent relationships.

The scatterplots of non-response models, Figure 6, illustrate the accuracy in fit of the developed models.

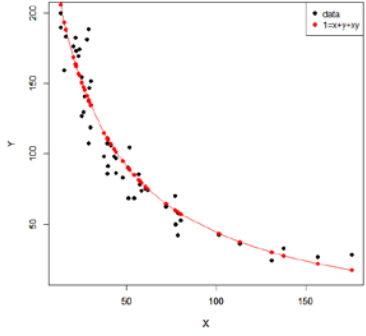 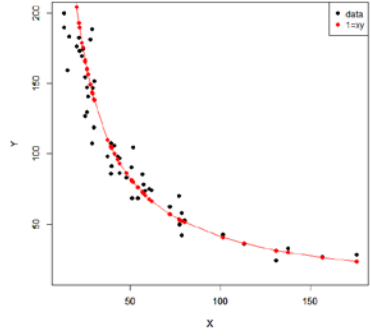

**Fig. 6a**  **Fig. 6b**

**Figure 6:** Scatter plot of non-response models; Fig. 6a illustrates the non-response model $1 = \alpha_1 x + \alpha_2 y + \alpha_3 xy$ and Fig. 6b illustrates the non-response model $1 = \alpha_1 xy$. These figures demonstrate the ability of non-response analysis to detect constants and codependent relationships.

Since the assumption of independence is not required, when solving for $\hat{x}$ and $\hat{y}$, the error terms are no longer perpendicular to the mean but rather is given by $\theta_T = \arccos\left(\frac{SSM+SSE-SST}{2\sqrt{SSM \times SSE}}\right)$. Consider the simulation with $\sigma = 5$, Table 2. In comparison to the measured angle of $90°$, the model with the best degree of separation is the non-response model including all the terms followed by the second-order standard regression model. In terms of the height, $h$, the distance between the point estimates and the line between the data and the means (Vos, 2010), the non-response model was the best followed by the second-order standard regression model.

**Table 2:** Estimations of Angles and Relative Heights

| Regression | Model | $\sigma = 1$ | | $\sigma = 5$ | |
|---|---|---|---|---|---|
| | | $\theta_T$ | $h$ | $\theta_T$ | $h$ |
| Rotational (Standard) | $y = \alpha_0 + \alpha_1 x + \alpha_2 xy$<br>$y = \alpha_0 + \alpha_1 x$ | 77.8 (6) | 33.3 (6) | 74.9 (3) | 34.880 (4) |
| Rotational (Standard) | $x = \alpha_0 + \alpha_1 y + \alpha_2 xy$<br>$x = \alpha_0 + \alpha_1 y$ | 68.3 (7) | 36.4 (7) | 65.2 (4) | 38.6 (6) |
| Rotational (Constant) | $xy = \alpha_0 + \alpha_1 x + \alpha_2 y$<br>$xy = \alpha_0$ | 81.8 (3) | 5.3 (2.5) | 52.2 (5) | 33.0 (3) |
| Standard | $y = \beta_0 + \beta_1 \frac{1}{x}$ | 77.9 (5) | 5.5 (4) | 12.8 (7) | 276.55 (7) |
| Standard | $y = \beta_0 + \beta_1 x + \beta_2 x^2$ | 87.4 (2) | 13.9 (5) | 87.5 (2) | 18.9 (2) |
| Non-Response | $1 = \alpha_1 x + \alpha_2 y + \alpha_3 xy$ | 89.1 (1) | 5.0 (1) | 89.9 (1) | 18.0 (1) |
| Non-Response | $1 = \alpha_1 xy$ | 80.9 (4) | 5.3 (2.5) | 48.9 (6) | 34.886 (5) |

## 2.2 Real World Data and Validation of Method

Throughout this paper's various simulations and interpretations regarding the concept of non-response and standard regression models, here is an application to that of real-world data, such as Boyle's law. Boyle's law (sometimes referred to as the Boyle–Mariotte law, or Mariotte's law), which is an experimental gas law that describes the product of the pressure and volume for a gas is a constant for a fixed amount of gas at a fixed temperature (Draper, 1861). Boyle's law uses inverse proportional rules, "where the absolute pressure exerted by a given mass of an ideal gas is inversely proportional to the volume it occupies if the temperature and amount of gas remain unchanged within a closed system" (Levine, 1978).

Consider the following three models: $1 = \alpha_1 x + \alpha_2 y + \alpha_3 xy$, $y = \beta_0 + \beta_1 x + \beta_2 x^2$, and $y = \beta_0 + \beta_1 \frac{1}{x}$. The concept of using inverse proportionality can be examined using non-response analysis in conjunction with standard regression to verify of Boyle's law. In mathematical terms, Boyle's law can be stated as

$$P \propto \frac{1}{V}$$

or

$$PV = k$$

where $P$ is the pressure of the gas, $V$ is the volume of the gas, and $k$ is a constant (Levine, 1978).

Now, consider Boyle's data (Boyle, 1662) (Fazio, 1992) measuring the *number of equal spaces in the shorter leg that contained the same parcel (parcal) of air diversely extended* represent the **volume** ($x$) and the aggregate of the *height of the mercurial cylinder in the longer leg that compressed the air into those dimension* and *the height of the mercurial cylinder that counter-balanced the pressure sustained by the included air* represent the **pressure** ($y$). With a coefficient of determination of 0.8595 and 0.8551, **volume** and **pressure** are comparable in variability, Figure 7; whereas with $R^2 = 0.9999878$, the produce of **volume** and **pressure** is extremely constant with small random error, Figure 7.

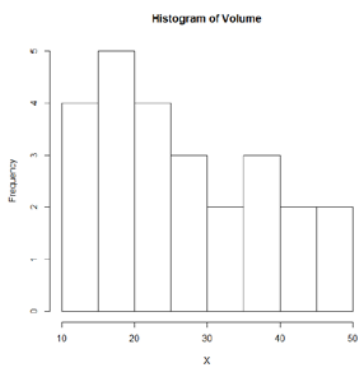 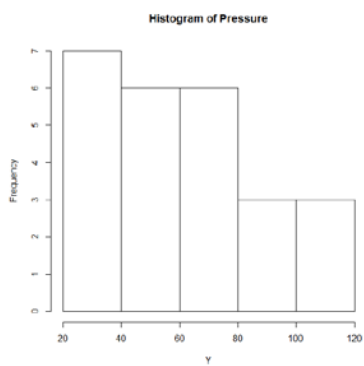 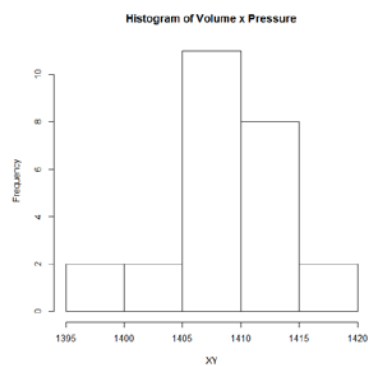

**Fig. 7a**          **Fig. 7b**          **Fig. 7c**

**Figure 7:** Histograms of variables; Fig. 7a is a depiction of the distribution of measured **volume**; Fig. 7b is a depiction of the distribution of measured **pressure**; and Fig. 7c depicts the distribution of the product of **volume** and **pressure**. The first two histograms reveal that they are not normally distributed; however the product of **volume** and **pressure** appears more symmetric and approximately normal.

As illustrated in Figure 8, the data from the controlled experiment shows minimal variance in both **volume** and **pressure**.

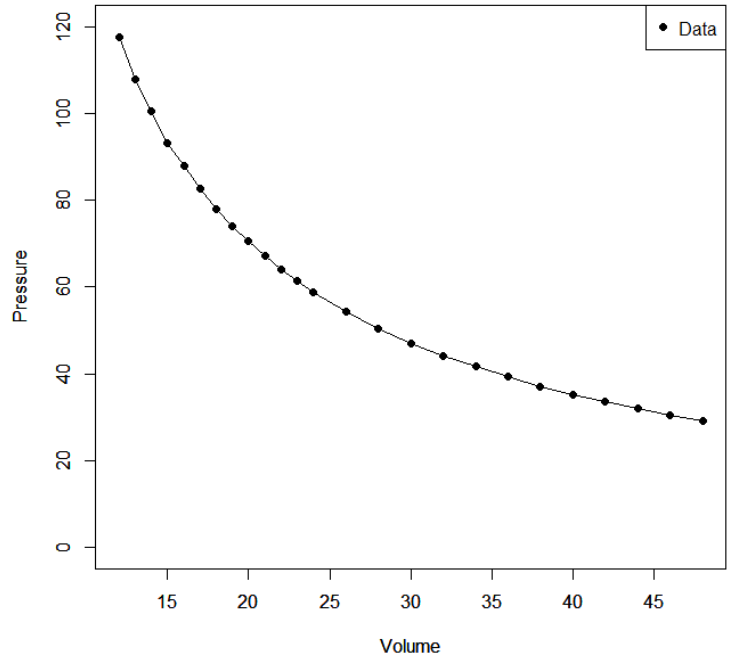

**Figure 8:** Scatter plot of **volume** and **pressure** observed by Boyle in 1662. This figure illustrates the inverse relationship that exist between **volume** and **pressure** with minimum random error.

In Figure 9, the fit of the three models depict that the nonresponse and inverse models are good fits with $\theta_T = 92.97, h = 0.01555$ and $\theta_T = 84.3, h = 0.02345$ respectively while the standard model was not a good fit with $\theta_T = 96.40, h = 0.94929$. It should be noted that when estimating volume, where complex solutions exist, the average or real part was used as the point estimate. This occurred for the first three points in standard second order model. The models outlined below including the associated degree of separation and the height (that is the extent to which the model estimates are removed from the data and their means, as they relate to the scatterplots in Figure 9 respectively.

$$1 = \alpha_1 x + \alpha_2 y + \alpha_3 xy; \, \theta_T = 92.97 \,; \, h = 0.01555$$
$$y = \beta_0 + \beta_1 x + \beta_2 x^2; \, \theta_T = 96.40; \, h = 0.94929$$
$$y = \beta_0 + \beta_1 \frac{1}{x}; \, \theta_T = 84.3; h = 0.02345$$

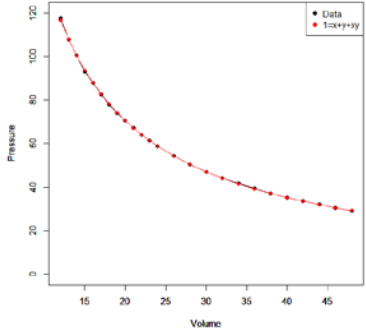 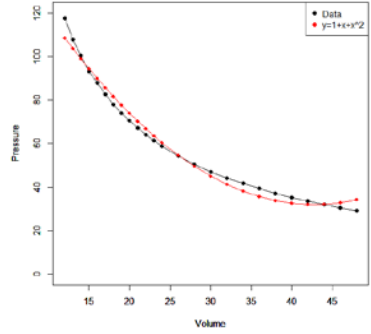 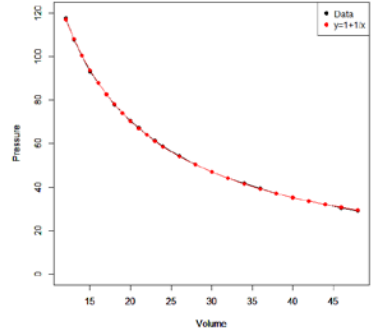

       Fig. 9a                        Fig. 9b                       Fig. 9c

**Figure 9:** Scatter plot of **volume** and **pressure** observed by Boyle and developed models; Fig. 9a illustrates the estimated pressure using the non-response model $1 = \alpha_1 x + \alpha_2 y + \alpha_3 xy$ and solving for $y$; Fig. 9b illustrates the estimated pressure using the standard regression model $y = \beta_0 + \beta_1 x + \beta_2 x^2$; and Fig. 9c illustrates the estimated pressure using the standard regression model $y = \beta_0 + \beta_1 \frac{1}{x}$.

## 3. Usefulness

Implicit regression analysis helps to identify and understand the constant nature of a variable and the interactive term in the model without any underlying distribution. Here, the problem of collinearity is addressed implicitly because the assumption of dependent and independent variables is not required as in classic regression model. The non-response model is the best fitting model.

Implicit regression analysis helps to better address multivariate random error in that it considers all relationships co-dependently; for example, the inverse relationship among measured variables have estimated random error in both directions. This analysis demonstrates and verifies Boyle's law using the data gathered on volume and pressure, and their interactions for the controlled experiment.

## 4. Discussion/Conclusion

In conclusion, it is recommended that researchers perform both rotational analyses and non-response analysis in conjunction with standard regression to tease out the dependence/independency relationships among variables.